\documentclass[times,twocolumn,final]{elsarticle}
\usepackage[margin=1.5cm]{geometry}
\usepackage{medima}
\usepackage{framed,multirow}

\usepackage{amssymb}
\usepackage{latexsym}

\usepackage[x11names]{xcolor}
\usepackage[
  colorlinks,
]{hyperref}

\AtBeginDocument{\hypersetup{citecolor=DodgerBlue4}}

\usepackage{verbatim}
\usepackage{bm}
\usepackage{amsmath}
\usepackage{setspace}
\usepackage{array}
\usepackage{multirow}
\usepackage{xcolor}

\begin{document}


\begin{frontmatter}

\title{Deep and Statistical Learning in Biomedical Imaging: State of the Art in 3D MRI Brain Tumor Segmentation}%

\author[1]{K. Ruwani M. Fernando \corref{cor1}}
\cortext[cor1]{Corresponding author.}
\ead{kfernando@usf.edu}
\author[1]{Chris P. Tsokos} 

\address[1]{Department of Mathematics and Statistics, University of South Florida, Tampa, FL, 33620,USA}


\begin{abstract}
Clinical diagnosis and treatment decisions rely upon the integration of patient-specific data with clinical reasoning. Cancer presents a unique context that influences treatment decisions, given its diverse forms of disease evolution. Biomedical imaging allows non-invasive assessment of diseases based on visual evaluations, leading to better clinical outcome prediction and therapeutic planning.  Early methods of brain cancer characterization predominantly relied upon the statistical modeling of neuroimaging data. Driven by breakthroughs in computer vision, deep learning has become the \emph{de facto} standard in medical imaging. Integrated statistical and deep learning methods have recently emerged as a new direction in the automation of medical practice unifying multi-disciplinary knowledge in medicine, statistics, and artificial intelligence. In this study, we critically review major statistical, deep learning, and probabilistic deep learning models and their applications in brain imaging research with a focus on MRI-based brain tumor segmentation. These results highlight that model-driven classical statistics and data-driven deep learning is a potent combination for developing automated systems in clinical oncology. 

\end{abstract}


\end{frontmatter}


\section{Introduction}
\label{intro}
Despite significant progress being made towards understanding the pathophysiology of cancer, its self-sustaining and highly dynamic nature poses several challenges pertaining to cancer detection and treatment monitoring.  Although technological advances hold promise in addressing these dilemmas, owing to their complexity, challenges remain in several stages of cancer management, including pre-cancerous lesion detection, tumor segmentation, tumor infiltration, tracing of tumor evolution, and prediction of tumor aggressiveness and recurrence \citep{bi2019artificial}. 

Brain tumors are among the most dangerous malignancies. The unique genetic, epigenetic, and microenvironmental properties of neural tissues confer resistance to treatment \citep{quail2017microenvironmental}, thus, the clinical management of brain tumors remains a critical challenge. Among tumors arising from the brain, the most predominant are metastases from systemic cancers and gliomas \citep{mcfaline2018brain}.  Stage IV glioma, referred to as glioblastoma, develops from glial cells and infiltrates the surrounding cell tissues. It is the most aggressive form of primary brain tumor, with a median survival of 15 months \citep{koshy2012improved}. Brain tumors are often complex in shape and vary greatly in size, texture, and location, thus, clinical information related to tumors exhibits high spatial and structural variation from patient to patient \citep{pereira2016brain}. 

\begin{figure*}[ht]
    \centering
    \includegraphics[width=\textwidth]{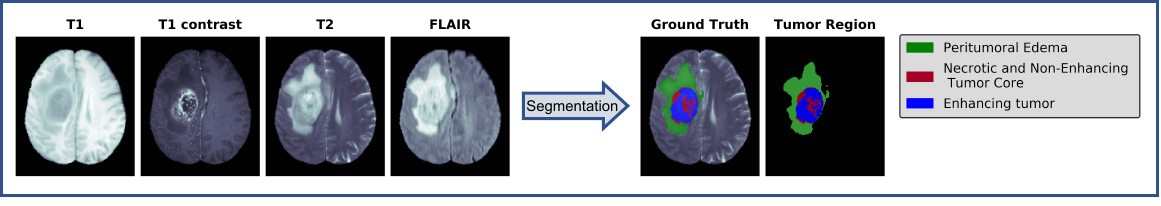}
    \caption{On the left is an example of a glioblastoma brain tumor in T1, T1-contrast, T2 and FLAIR modalities. On the right is the ground truth segmentation of the tumor that specifies the class of each voxel.}
    \label{fig:gbm}
\end{figure*}

Neuroimaging is a pivotal diagnostic tool in disease assessment. While a variety of diagnostic imaging techniques such as X-ray, computed tomography (CT), positron emission tomography (PET), and mammograms are currently in use, magnetic resonance imaging (MRI) is the method of choice in neurology as it does not involve the risk of exposure to ionizing radiation \citep{bhargava2013contrast}. In brain tumor studies, the clinical information embedded in MRI data is vital for analyzing structural brain abnormalities, detecting pre-cancerous lesions, tumor segmentation, and diagnosis.

Segmentation of biomedical images entails the delineation of pathological regions. In the case of glioblastoma multiforme brain tumor (GBM), segmentation involves separating different heterogeneous regions of tumor tissues as the necrotic core, enhancing tumor, and peritumoral edema \citep{bakas2018identifying}. Different types of imaging modalities provide useful biological information related to tumor-induced neural tissue changes. T1-weighted (T1), T1-weighted contrast-enhanced (T1ce), T2-weighted (T2), and T2 fluid-attenuated inversion recovery (FLAIR) modalities are the routinely used brain MRI image sequences \citep{shukla2017advanced}. A 2D slice of each modality and the ground truth segmentation map are depicted in Fig. \ref{fig:gbm}. 



Traditional methods of tumor evaluation depend on qualitative features such as tumor density, intra-tumoral cellular composition, and anatomic relationship to surrounding tissues, also termed semantic features. Technological advancements have enabled the quantitative assessment of neuropathology, facilitating the quantification of tumor size, shape, and textual patterns \citep{aerts2014decoding}. While early methods of cancer diagnosis relied upon probabilistic tumor models, pivotal breakthroughs in computer vision have led to a rapid rise in Artificial Intelligence (AI) research in medical imaging \citep{bi2019artificial}. Deep learning \citep{lecun2015deep} is a subset of AI that has great potential to support clinicians in gaining insight from complex, high-dimensional, and heterogeneous biomedical data. Traditional statistics and deep learning vary in their approaches to brain tumor quantification, each with its own strengths and weaknesses. Although biomedical image segmentation with statistical modeling and deep learning methods are well-established as independent areas of research, hybridization of the two paradigms is an emerging trend. Going beyond recent retrospective reviews on tumor segmentation which are mostly concentrated only on deep learning-based methods, we provide a comprehensive overview linking the two disciplines. 

Our \emph{contributions} can be summarized as follows:

\begin{itemize}
\item 
The main contribution of this study towards neuroimaging literature is to provide an integrated overview of statistical, deep learning, and probabilistic deep learning methodologies for brain tumor segmentation from both theory-driven and application perspectives.

\item To the best of our knowledge, no previous review studies link the two disciplines of statistics and deep learning in biomedical segmentation. To fill this research gap, we provide an overview of probabilistic deep learning with a structured taxonomy of different applications of hybrid statistical and deep learning methods in brain tumor segmentation. 

\item Through a comprehensive summary of statistical and deep learning literature, we provide a critical discussion related to the methodologies, limitations, key challenges, and future directions in the area of brain tumor segmentation.
\end{itemize}

The remainder of this paper is structured as follows: In Section \ref{sec:preli}, we present conceptual bases pertaining to cortical segmentation. Section \ref{sec:stat} provides a statistical overview comprised of computational theories and recent studies in 3D MRI brain tumor segmentation. This is followed in Section \ref{sec:DL} by deep learning network architectures for medical imaging and their tumor segmentation applications. Section \ref{sec:PDL} presents a concise overview of integrated statistical and probabilistic deep learning studies in neuroimaging. Section \ref{sec:data} details the data sets and evaluation metrics related to brain tumor segmentation. Section \ref{sec:limitations} provides a discussion on limitations, critical challenges, and future directions. Finally, Section \ref{sec:conclude} concludes the paper.


\section{Brain Tumor Segmentation Preliminaries}
\label{sec:preli}

Brain imaging analyses rely upon anatomical images of the brain generated using an MRI scanner. The basic concepts of the tumor segmentation process and MRI scan pre-processing are briefly described here.

\subsection{Tumor Segmentation and its clinical applications}

Segmentation is the process of partitioning an image into a set of semantically (i.e., anatomically) meaningful and coherent regions, the result of which is either the contours that delineate the region boundaries or the homogeneous region of the structures of interest (e.g., tumors) \citep{despotovic2015mri}. 

A 3D image can be defined as a continuous function $f(s,t,u)$, such that $f \colon \mathbb{R}^3 \to \mathbb{R}$, where $f(s,t,u)\in \mathbb{R}$ represents the image intensity level at spatial coordinates $(s,t,u) \in \mathbb{R}^3$. Let the image domain be denoted as $\mathcal{I}$ and the segmentation classes by $k \in K$, where each voxel $i$ belongs to one of the $K$ possible classes such that the set $S_k \subset \mathcal{I}$. Then the segmentation task involves determining the set $S_k$ that satisfies:
\begin{equation}
    \mathcal{I} = \bigcup_{k=1}^K S_k
\end{equation}
where $S_j \cap S_k = \emptyset$ for $j \neq k.$

In the context of medical imaging, segmentation involves identifying pixels or voxels that represent organs, lesions, and other anatomical substructures. Thus, it enables the identification of the extent of an abnormality which can range from the estimation of the maximal tumor diameter in 2D measurements to whole tumor assessments and adjacent tissue evaluations in 3D volumetric segmentation. Such information can be subsequently applied in diagnosis, risk assessment, and therapeutic planning. 

There are several clinical applications of cortical segmentation \citep{suri2001advanced}, some of which are summarized here. In neurosurgery, \textit{surgical planning} is imperative, and an accurate segmentation can be utilized to identify the 3D coordinates of tumor cells. Segmentation also aids in \textit{neuro-surgical navigation} to guide the neurosurgeon to move towards anatomical targets such as tumors while avoiding critical regions of the brain. \textit{Quantitative assessment} of pathology is vital for establishing treatment strategies and often necessitates segmentation as it allows quantitative assessment of the extent of tumor resection and measurement of tumor volume. By separating anatomical structures in image data into spatially and structurally correlated regions, segmentation enables the effective \textit{visualization} of complex biomedical images.

\subsection{MRI pre-processing in tumor segmentation}

In structural brain MRI analysis, certain pre-processing steps are applied prior to automated segmentation which typically includes image registration, skull stripping, and bias field correction. 

\begin{itemize}

\item  \textit{Image registration} \citep{zitova2003image} involves spatially aligning images acquired by different sensors, at different times, from different viewpoints to a common anatomical space. Inter-patient registration aims to standardize images across subjects, whereas intra-patient registration aids in the alignment of imaging modalities (e.g., T1 and T2) within the same subject. The accuracy of inter-subject glioma registration with linear and non-linear transformations was studied in \citep{visser2020accurate}. They demonstrated that a linear transformation is sufficient to register the MRI images to an anatomical reference space and to capture the distribution of tumor locations within a patient group.  

\item \textit{Skull stripping} \citep{kalavathi2016methods} is the process of brain tissue extraction from extra-meningeal tissues such as the skull, muscles, or fat and aims at classifying voxels as brain or non-brain. The most common approach for skull stripping in early studies is the use of tools such as the brain extraction tool (BET) \citep{smith2002fast}. Nevertheless, early methods such as BET were designed for non-pathologically affected brains, and it has been observed that they do not work appropriately in brain tumor scans. Skull stripping methods have now been designed and developed that specifically focus on brain tumor scans \citep{thakur2019skull}.

\item Intensity inhomogeneity, also known as the \textit{bias field} \citep{belaroussi2006intensity}, is an undesirable artifact that degrades segmentation performance leading to quantitative and qualitative misinterpretations. Intensity non-uniformity can be characterized by the intensity variation of the same tissue across the MR image. To eliminate the spatial inhomogeneities, several methods have been proposed \citep{belaroussi2006intensity, song2017review}, and N4 bias field correction \citep{tustison2010n4itk} is one of the widely applied methods. However, prior literature also suggests that bias field correction yields inferior results when applied as a pre-processing step as it obliterated the T2-FLAIR signal \citep{bakas2017advancing}.
\end{itemize}

\begin{figure*}[ht!]
    \centering
    \includegraphics[width=\textwidth]{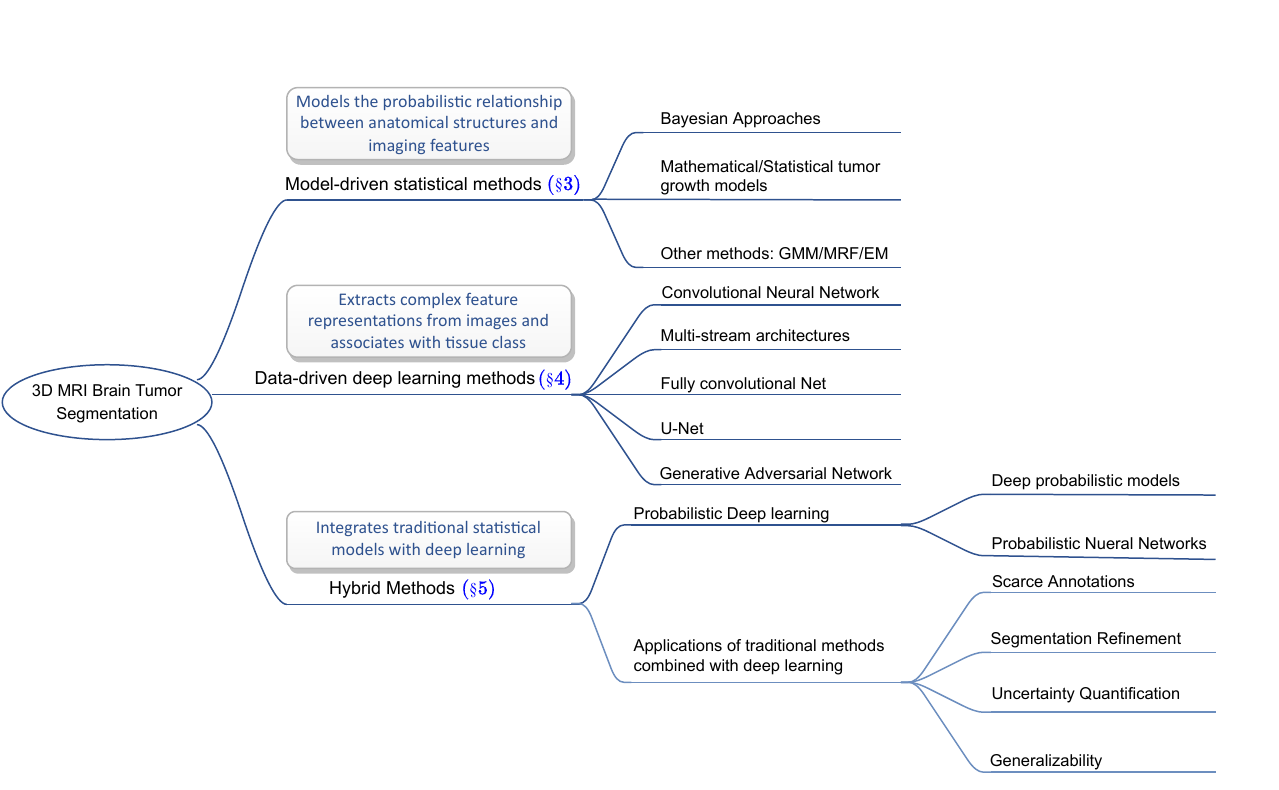}
    \caption{Flow diagram of tumor segmentation methods: Statistical (section \ref{sec:stat}), Deep-learning (section \ref{sec:DL}), and integrated deep and statistical methods (section \ref{sec:PDL}) with their applications (sub-section \ref{subsec:applications}).
    \label{fig:review_methods}}
\end{figure*}

The current clinical practice in oncology is typically dependent on manually traced segmentation which is time-consuming and subject to inter-observer variability leading to inconsistent segmentation and poor reproducibility. Thus, a time-efficient, generalizable and reproducible automated segmentation process is highly desirable. Several different segmentation techniques have been proposed in the past literature, including intensity-based, atlas-based, and surface-based methods, but here we primarily focus on statistical, deep learning, and probabilistic deep learning based segmentation. The rest of the paper is structured as depicted in Fig. \ref{fig:review_methods}.


\section{Model-driven statistical methods}
\label{sec:stat}

Prior studies in tumor segmentation of medical images often relied upon statistical inference. Statistical and probabilistic modeling of brain tumors typically involves the incorporation of domain-specific prior knowledge related to the anatomy, appearance, and spatial distribution of the tissues. For instance, the spatial distribution prior for a given set of 3D MRI images can be obtained via lesion-specific bio-markers or by employing a tumor growth model (sub-section \ref{subsec:tumor growth}) that can be used to deduce the most probable tumor localization. While these methods generalize well to unseen images, encoding prior information for a tumor is challenging. Here we present some of the common statistical approaches used in tumor segmentation. 




\subsection{The Bayesian Modeling Framework}

The Bayesian approach can be used to model the probabilistic relationship between tissue class and imaging features such as voxel intensity values \citep{van2015tissue}. In the Bayesian framework, the model is fully specified by the prior and likelihood distributions. 

The prior indicates the current state of the knowledge with regard to likely configurations of anatomical structures in the image before the data is observed. Let the segmentation prior distribution be denoted by $p(\mathbf{y}|\bm{\theta}_y)$ where $\mathbf{y}=(y_1,\dots,y_I )^T$ are the labels in an image with $I$ voxels such that $y_i \in \{1,\dots,K\}$ represents one of the $K$ possible anatomical structures assigned to voxel $i$ and $\bm{\theta}_y$  is the set of parameters of prior. The likelihood is the conditional distribution $p(\mathbf{x}|\mathbf{y},\bm{\theta}_x)$ of the voxel intensities $\mathbf{x}=(x_1,\dots,x_I )^T$ given anatomical labels $\mathbf{y}$ where $\bm{\theta}_x$  is the set of parameters. The segmentation posterior can then be computed using the Bayes’ rule with the appropriate parameters $\hat{\bm{\theta}}$: 

\begin{equation}
    p(\mathbf{y}|\mathbf{x}, \hat{\bm{\theta}}) = \frac{p(\mathbf{x}|\mathbf{y},\hat{\bm{\theta}}_x) \, p(\mathbf{y}|\hat{\bm{\theta}}_y)}{p(\mathbf{x}|\hat{\bm{\theta}})}
\end{equation}
where $p(\mathbf{x}|\mathbf{y},\hat{\bm{\theta}}_x)$ is the likelihood, $p(\mathbf{y}|\hat{\bm{\theta}}_y)$ is the prior probability distribution and $p(\mathbf{x}|\hat{\bm{\theta}})$ is the marginal distribution with $p(\mathbf{x}|\hat{\bm{\theta}})= p(\mathbf{x}|\mathbf{y},\hat{\bm{\theta}}_x) \, p(\mathbf{y}|\hat{\bm{\theta}}_y).$

Prior and likelihood parametric distributions depend on parameters $\bm{\theta}=({\bm{\theta}_y}^T,{\bm{\theta}_x}^T)^T$, which are either assumed a priori known or estimated from the image data. The segmentation map of a new image can be estimated via maximum a posteriori (MAP) estimation which involves maximizing the posterior distribution $p(\mathbf{y}|\mathbf{x}, \hat{\bm{\theta}})$ of possible segmentations:
\begin{align} \label{eq1}
\hat{\mathbf{y}} & = \operatorname*{arg\,max}_\mathbf{y} p(\mathbf{y}|\mathbf{x}, \hat{\bm{\theta}}) \nonumber \\
 & = \operatorname*{arg\,max}_\mathbf{y} p(\mathbf{x}|\mathbf{y},\hat{\bm{\theta}}_x) \, p(\mathbf{y}|\hat{\bm{\theta}}_y)
\end{align}
where $p(\mathbf{x}|\hat{\bm{\theta}})$ is a constant that can be excluded.

Early studies, as well as current work, investigate the applicability of the Bayesian framework in brain MRI analyses \citep{corso2008efficient, raju2018bayesian, raja2020brain, narasimha2021effective, wang2022adaptive}. A Bayesian method for computing model-aware affinities was presented in \citep{corso2008efficient}, which has been shown to improve the segmentation of glioblastoma multiforme brain tumors. In \citep{raju2018bayesian}, the authors proposed a method for brain tumor segmentation and classification, in which they used a Bayesian fuzzy clustering approach to segment brain tumors. Multiple other studies in the previous literature have explored the applicability of fuzzy Bayesian-based segmentation in brain tumor detection \citep{raja2020brain, narasimha2021effective}. In a recent work \citep{wang2022adaptive}, the authors present an adaptive sparse Bayesian model combined with a Markov random field and Gibbs random field for brain tumor segmentation. Their experimental results indicate superior segmentation performance compared to other methods in terms of Dice coefficient and sensitivity. Although Bayesian MRI segmentation is a well-studied problem, it necessitates the use of prior information, and the analysis is sensitive to the choice of the prior. When encoding prior knowledge, the robustness of the prior must be ensured such that the conclusions are not affected by the form of the prior distribution. 


\subsection{Biophysical/mathematical tumor growth models in segmentation of brain tumors }
\label{subsec:tumor growth}

The characterization of dynamic biological processes through mathematics, modeling, and simulation has been extensively explored over the past decade \citep{rockne20192019}. Biophysical models simulate physiological and biological systems through mathematical formulations of the underlying processes. In cancer research, biophysical tumor growth models facilitate the quantification of tumor volume changes over time. Such models can be used to predict the development and progression of a tumor, enabling personalized treatment.

A substantial body of previous literature has studied mathematical and statistical tumor growth models that aid in tumor segmentation \citep{gooya2012glistr, bakas2015glistrboost, pei2017improved,lipkova2019personalized,mang2020integrated, sasank2022automatic}. In a review conducted in \citep{mang2020integrated}, the authors emphasize that identifying quantifiable tumor growth patterns through biophysical modeling leads to improved detection and segmentation of tumor sub-regions. The Glioma Image Segmentation and Registration (GLISTR) framework proposed in \citep{gooya2012glistr} attempts to solve a joint registration and segmentation problem based on a glioma growth model. The method proposed in \citep{bakas2015glistrboost} is an extension of the approach presented in \citep{gooya2012glistr}, in which tumor growth modeling is combined with gradient boosting for the segmentation of gliomas. In \citep{pei2017improved}, a tumor growth model is used to generate cell density patterns which are then utilized to improve tumor segmentation. In a recent study \citep{lipkova2019personalized}, the authors integrated structural and metabolic scans and modeled the tumor density of patients with glioblastoma  using a tumor growth model under the Bayesian framework. The Bayesian approach enables quantifying the tumor and imaging uncertainties, which are then propagated through the tumor models to obtain patient-specific tumor cell densities. The Bernoulli distribution is used to model the probability of a tumor observation with a simulated tumor cell density. 

Tumor dynamics vary significantly between patients, as well as across cancer cells in the same patient. Modeling tumor progression, therefore, is a complex process that is challenging to summarize mathematically.



\subsection{Other statistical approaches used in tumor segmentation}

\subsubsection{Gaussian Mixture Model} In the segmentation of brain MRI, voxel intensities are typically assumed to be independent samples from a mixture of Gaussian distributions. The Gaussian mixture model (GMM) assumes that the occurrence of a certain anatomical label in a particular voxel does not depend on the labels in other voxels. There have been several attempts to integrate GMMs in tumor segmentation \citep{chaddad2015automated,xia2016brain,byale2018automatic,chen2020rfdcr}. To extract imaging features from MRI scans of patients diagnosed with Glioblastoma, Ahmad et al. \citep{chaddad2015automated} used a GMM, which they demonstrated performed best compared to principal component analysis (PCA) and wavelet-based features. In \citep{xia2016brain}, a GMM based on variational Bayesian inference for brain MRI image segmentation was proposed. However, their approach requires training a large number of variational models, which is computationally expensive. A segmentation and classification approach for brain tumors was presented in \citep{byale2018automatic}, in which the segmentation was carried out using a GMM. In another study \citep{chen2020rfdcr}, the authors use a cascaded random forest together with a conditional random field for brain lesion segmentation, where they use a GMM to extract contextual features to feed into the classifier as input. 


The assumption of spatial independence between voxel intensities is not justifiable and may lead to voxel misclassification. Therefore, probabilistic tumor models should incorporate the context and characterize the true shape of the anatomical structures.

\subsubsection{Markov Random Field}

A voxel intensity value is statistically dependent on  the  intensity of neighboring voxels. Markov random field (MRF) allows modeling the spatial context by integrating voxel dependencies in the image surface. An MRF can be characterized by a graph $\mathcal{G} \triangleq (\mathcal{H} ,\mathcal{N})$ with $\mathcal{H}$ representing a lattice containing nodes and $\mathcal{N}$ indicating the links between the nodes, such that the neighboring relationship of nodes is defined by  $\mathcal{N}=\{\mathcal{N}_i | \forall i \in \mathcal{H} \}$, where $\mathcal{N}_i$ is the neighborhood around a position $i$ \citep{li2009mathematical}. This graphical structure is comparable to a 3D image grid with nodes representing the voxels and edges, the spatial dependency between voxels.

The MRF model reduces the effect of statistical noise in highly noisy data and the potential number of misclassified voxels. It has been successfully applied in many studies to segment brain tumors \citep{capelle2000unsupervised, gering2002recognizing, bauer2011segmentation, rajinikanth2017firefly, shahvaran2021morphological, barzegar2022efficient}. In \citep{capelle2000unsupervised}, the authors proposed an unsupervised MRF-based segmentation approach to detect the presence of tumors. They applied MRF to encode neighborhood dependency of pixels, where they then estimated the realization of the hidden Markov field via a maximum a posteriori estimator (MAP). Gering et al. \citep{gering2002recognizing} presented a brain tumor segmentation framework where they first trained on healthy brains to diagnose brain tissue abnormalities and then used a multi-layer MRF framework combined with extended expectation maximization (EM)-based segmentation. They incorporated contextual information at multiple levels in the segmentation process. However, this approach has certain limitations as high-level layer classification is sensitive to the noise in the lower layers. Bauer et al. \citep{bauer2011segmentation} utilized an MRF-based tumor growth model along with atlas registration for volumetric MRI brain tumor segmentation. To ensure that the atlas and MRI scan of the patient match, the tumor growth model is devised as a mesh-free MRF energy minimization problem. Their method is non-parametric and can be applied for tumor mass effect simulations which facilitate the atlas-based segmentation. Rajinikanth et al. \citep{rajinikanth2017firefly} applied the MRF-EM combination in conjunction with a firefly-assisted algorithm for MRI tumor segmentation. During the segmentation process, MRF assigns the label of each pixel to one of the three categories: white matter, gray matter, and tumor mass. Their results suggest that the proposed approach is a better alternative compared to the other methods tested in their study. In another study \citep{shahvaran2021morphological}, a morphological region-based active contour model was proposed for brain tumor extraction, in which an MRF was used to model the correlation between neighboring pixels. In a more recent study \citep{barzegar2022efficient}, glioma segmentation is formulated as an MRF-based energy optimization problem. 

While the MRF model allows integrating prior that incorporate spatial context, computing the posterior explicitly is computationally expensive and requires the use of approximation schemes. Thus, MRFs have shortcomings with respect to computational efficiency.

\subsubsection{Expectation Maximization Algorithm}

Expectation Maximization (EM) is a parameter optimization algorithm typically used in conjunction with other models during segmentation. The EM framework allows obtaining maximum likelihood estimates of unknown parameters, which is a common strategy in tumor segmentation when modeling MRI intensities of brain tissues as Gaussian mixtures. The EM algorithm iterates between two steps until convergence: the \emph{expectation step} (e-step) and the \emph{maximization step} (m-step). In brain tissue segmentation, the model parameter estimates are often initialized via a GMM. Then, in the E-step, the model seeks to estimate the tumor segmentation where each pixel/voxel is assigned to a cluster under the current model parameter estimates of the posterior distribution, and the M-step estimates the model parameters based on the current classification.

There is a considerable body of literature on tumor segmentation that has integrated the EM algorithm in conjunction with the hidden Markov random field (HMRF) model \citep{zhang2001segmentation, nie2009automated, doyle2013fully}. In \citep{zhang2001segmentation}, the authors showed that a fully automated 3D brain MR image segmentation that is both accurate and robust can be achieved through an HMRF-EM framework. The HMRF framework aids in encoding spatial information of neighboring sites, and the model parameters are estimated via maximum likelihood estimation using the EM algorithm. Moon et al. \citep{moon2002model} proposed a model-based approach for segmenting healthy and tumor brain tissues from 3D MRI, where they used a probabilistic geometric model coupled with the EM algorithm. They extended the spatial prior of a probabilistic normal human brain atlas with tumor information obtained from patient data to include spatial prior for tumor tissues. A generalized EM segmenter via a generative probabilistic modeling approach for brain lesion segmentation was introduced by Menze et al. \citep{menze2015generative} that demonstrated improved generalization. The Brain Tumor Segmentation (BRATS) challenge \citep{menze2014multimodal, bakas2017advancing, bakas2018identifying} primarily focuses on the segmentation of brain tumors and the winning method of BRATS2015 \citep{bakas2015glistrboost} and the third place in BRATS2016 \citep{zeng2016segmentation} were also based on EM. Similar work has also been pursued by others in recent studies \citep{qiao2019data,bhima2022contemporary}, where the EM algorithm has been incorporated in the segmentation of brain tumors. 

The EM algorithm ensures that likelihood increases with each step and is frequently used when the model depends on latent variables. Nevertheless, it exhibits slow convergence and may only converge to local optima. 

\begin{table*}[ht!]
\footnotesize
\caption{Overview of brain tumor detection studies based on statistical methods}
\label{tab:stat}
\renewcommand{\arraystretch}{1.5} 
\resizebox{\textwidth}{!}{\begin{tabular}{>{\raggedright\arraybackslash}p{3cm} c c p{8.5cm}}
\hline
\multicolumn{1}{c}{\textbf{Method}} & \multicolumn{1}{c}{\textbf{Reference}} & 
\textbf{Year} & \multicolumn{1}{c}{\textbf{Remarks}}  \\ 
\hline
Bayesian Modeling
& \citep{corso2008efficient} & 2008 & Incorporate model-aware affinities calculated through a Bayesian approach.\\
& \citep{raju2018bayesian} & 2018  &  Bayesian fuzzy clustering-based tumor segmentation.\\
& \citep{raja2020brain} & 2020  &  Bayesian fuzzy clustering-based tumor segmentation.\\
& \citep{narasimha2021effective} & 2021  & Fuzzy Bayesian segmentation approach, in which fuzzy and Gaussian naive Bayes strategies are incorporated in the tumor cut algorithm. \\
& \citep{wang2022adaptive} & 2022 & Sparse Bayesian modeling approach combined with a Markov random field and Gibbs random field.\\
Biophysical/mathematical tumor growth modeling
& \citep{gooya2012glistr} & 2012 & A glioma growth model is used to solve a joint registration and segmentation problem.\\
& \citep{bakas2015glistrboost} & 2015 & EM-based tumor growth modeling combined with gradient boosting. The winning method of BRATS2015 challenge.\\
& \citep{pei2017improved} &
2017 & A tumor growth model is used to generate cell density patterns which are then utilized to improve tumor segmentation.\\
& \citep{lipkova2019personalized} & 2019 &
Modeled the tumor density of patients with glioblastoma using a tumor growth model under the Bayesian framework.\\
& \citep{sasank2022automatic}  & 2022 & Tumor cell density patterns are extracted using a tumor growth model, which are then fed as input to a convolutional network for segmentation.\\
Gaussian Mixture Models (GMM)
& \citep{chaddad2015automated} & 2015 & GMM-based glioblastoma feature extraction approach.\\      
& \citep{xia2016brain} & 2016 & A GMM based on variational Bayesian inference.\\ 
& \citep{byale2018automatic} & 2018 & The tumor region was segmented through a GMM.\\
& \citep{chen2020rfdcr} & 2020 & A GMM was used to extract contextual features from MRI scans.\\
Markov Random Field (MRF)
&  \citep{capelle2000unsupervised} &  2000   & The neighborhood dependency of pixels in MRI scans was encoded though an MRF. \\ 
& \citep{gering2002recognizing} & 2002 & A multi-layer MRF combined with extended EM-based segmentation approach. \\
& \citep{bauer2011segmentation} & 2011 & Utilized an MRF-based tumor growth model together with atlas registration.\\
& \citep{rajinikanth2017firefly} & 2017 & Applied the MRF-EM combination in conjunction with a firefly-assisted algorithm.\\
& \citep{shahvaran2021morphological} & 2021 & The correlation between neighboring pixels was modeled by an MRF.\\
& \citep{barzegar2022efficient}   & 2022 & Glioma segmentation is formulated as a MRF energy optimization problem.\\
Expectation Maximization (EM)
&  \citep{zhang2001segmentation} &  2001   & HMRF framework is utilized to encode spatial information of neighboring sites and the model parameters are estimated via maximum likelihood estimation using the EM algorithm. \\ 
& \citep{moon2002model} & 2002 & Used a probabilistic geometric model coupled with the EM algorithm.\\
& \citep{menze2015generative} & 2015 & Proposed a generalized EM segmenter via a generative probabilistic modeling approach which demonstrated improved generalization. \\
& \citep{zeng2016segmentation} & 2016 &  A hybrid generative-discriminative model. The third place in BRATS2016 challenge.\\
& \citep{qiao2019data} & 2019 &  Evaluate GMM-EM algorithm for its applicability in lesion segmentation.\\
&\citep{bhima2022contemporary} & 2022 &  Used a probabilistic EM-GMM approach to improve segmentation accuracy.\\
\hline
\end{tabular}}
\end{table*}

\subsection{Summary} 
In this section, we reviewed the statistical approaches used for modeling and optimization of brain tumor segmentation models, summarised in Table \ref{tab:stat}. Justifying the underlying probability distribution that drives the data is an important facet of statistical model-based segmentation methods. These methods have the intrinsic limitations of requiring domain expertise and knowledge of prior distribution. Translating information about prior knowledge into a probabilistic model can be challenging, and inference may involve extensive computations \citep{icsin2016review}. Furthermore, statistical tumor models are often parameterized by many parameters. Estimation of unknown parameters could be computationally challenging due to the difficulties posed by mathematical issues such as non-convexity and model uncertainties. These shortcomings restrict their deployment in clinical practice. However, model-based approaches are effective in terms of model interpretability and the amount of data required for the analysis.

\section{Data-driven deep learning based methods}
\label{sec:DL}

Deep Learning (DL) excels at extracting complex feature representations that the human brain cannot perceive, thereby enabling quantifiable image interpretation. Previous studies have exploited different architectures of deep networks for biomedical image segmentation, varying from unique convolutional neural network based architectures to probabilistic deep generative models. In this section, we present widely adopted deep learning architectures in the recent literature and their practical applications in 3D multi-modal MRI brain tumor segmentation.

\subsection{Convolutional Neural Networks}

Recent years have seen a proliferating use of convolutional neural networks (CNNs) \citep{lecun1999object} in medical image analysis owing to their ability to integrate structural information contained in adjacent pixels. CNNs are comprised of multiple convolutional and pooling layers that are stacked sequentially. By convolving a filter (also referred to as a kernel) across the input and performing a convolution operation, a convolutional layer generates feature maps, which are then passed through an activation function such as a Rectified Linear Unit (ReLU) for non-linear transformation. The output feature map is then fed to a pooling layer for down-sampling. The fully connected layer placed at the end of the network maps the features extracted by convolutional and pooling layers into the final output and makes the final prediction. 

\begin{figure*}
    \centering
    \includegraphics[width=0.95\textwidth]{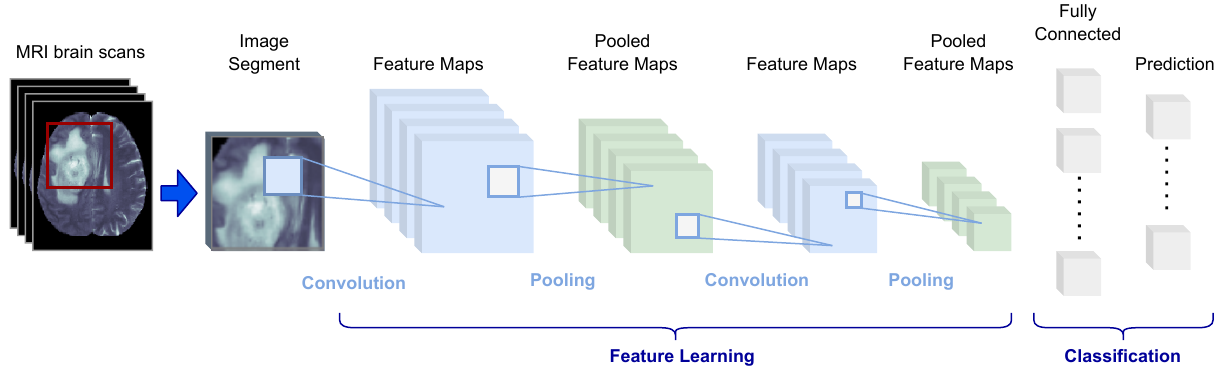}
    \caption{Patch-wise segmentation with CNN architecture. The input to the CNN is an image patch and the output is the probabilities for each class where the prediction for center pixel is the class with the highest score.}
    \label{fig:cnn}
\end{figure*}

In contrast to conventional DL classifiers such as AlexNet \citep{krizhevsky2017imagenet}, VGGNet \citep{simonyan2014very} and GoogLeNet \citep{szegedy2015going}, which take an image as the input and output a probability distribution over all classes with regard to the entire image, the task of segmentation requires a per-pixel classification of the input. This can be achieved by feeding the classifier with patches extracted around each pixel and adopting a sliding-window strategy (Fig. \ref{fig:cnn}). Semantic segmentation in the computer vision domain typically involves 2D images, consequently, the methods developed for segmentation are primarily based on 2D methods. Therefore, deep learning in 3D volumetric medical image segmentation requires modifications to the architectures developed for semantic segmentation. While some studies have shown promising results when applying 2D CNN architectures to process 3D volumes in a slice-by-slice manner, for example in \citep{zikic2014segmentation,dvovrak2015local,pereira2016brain, havaei2017brain}, it is arguably a sub-optimal use of 3D clinical image data as it is time-inefficient and fails to leverage contextual information from adjacent slices. 

\emph{3D Convolutional Neural Networks} (3D CNNs), as a generic extension of 2D CNNs capable of extracting volumetric representations with the use of 3D kernels have received significant attention in neuroimaging. In the seminal work of Kamnitsas et al. \citep{kamnitsas2017efficient}, a 3D CNN-based architecture, termed \emph{DeepMedic}, was introduced for automated brain lesion segmentation of 3D volumetric brain scans and ranked at the top in the Ischemic Stroke Lesion Segmentation (ISLES 2015) challenge. They conducted a patch-based analysis using a dual pathway strategy which allows feeding the network a version of the image at multiple scales concurrently, facilitating the incorporation of both local and global contextual information. This process was further improved later with several architectural modifications \citep{kamnitsas2017ensembles}. 

Patch-wise segmentation with CNNs does not require major architectural adaptations. However, it is computationally expensive and entails redundant operations as adjacent pixels in an image are similar in the vicinity of the pixel being processed, and the input patches from adjacent pixels overlap \citep{haque2020deep}. In addition, the smaller the patch size, the less likely the network is to learn the global context as the learning process is confined to the visible features in each patch.

\subsection{Multi-path architectures} 

The integration of contextual 3D information can also be performed through multi-stream architectures. While the standard CNN can incorporate multiple sources of information through the channels in the input layer, multi-stream architectures enable the merging of the channels at intermediate points in the network. They enable fusing features extracted from multi-angled patches from the 3D space (three patches from the orthogonal views of an MRI volume), an approach known as the 2.5D method \citep{roth2014new}, which incorporates (partial) 3D information from neighboring slices using 2D kernels that results in lower computational cost, but less powerful volumetric representations than in 3D CNN. The multi-path approach also facilitates combining patches of different scales (2D, 2.5D, 3D patches), where each branch contains a CNN corresponding to each patch type. The idea of multi-path networks has been successfully implemented by several authors in the context of medical imaging \citep{zhao2016multiscale,havaei2017brain, kamnitsas2017efficient, razzak2018efficient, sun2021segmentation}. However, multi-stage training schemes can be susceptible to over-fitting.



\subsection{Fully Convolutional Net} 

\begin{figure*}[ht!]
    \centering
    \includegraphics{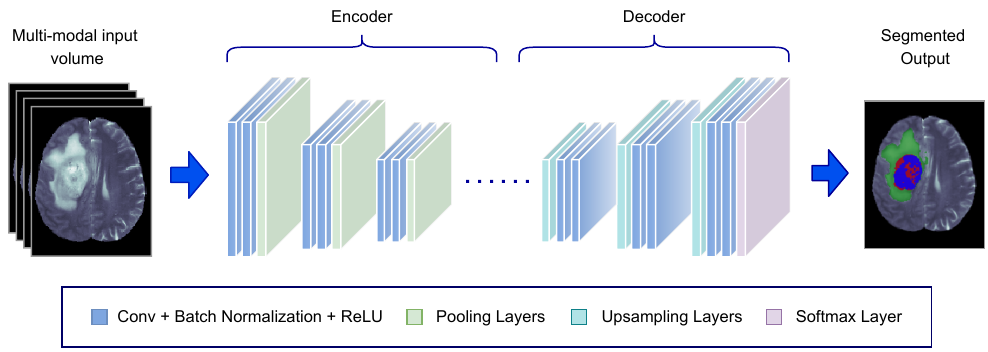}
    \caption{A schematic illustration of semantic-wise brain tumor segmentation with a SegNet-like architecture. The down-sampling through pooling layers in the encoder part of the network is compensated by the up-sampling layers in the decoder part that results in a segmentation output with dimensions same as the original input image.}
    \label{fig:seg}
\end{figure*}

To address the limitations of CNNs, Long et al. \citep{long2015fully} devised a fully convolutional network (FCN) by transforming fully connected layers in the standard CNN with convolutional layers that enable taking an input of arbitrary size and generating correspondingly sized output. While the FCN can be efficiently applied to an image or volume, the use of convolution layers with filters larger than 1 x 1 and max-pooling layers or strided convolutions leads to a reduction in the data dimension resulting in a very low resolution of the output compared to the input. To circumvent this limitation, the authors in \citep{long2015fully} applied the concept of an up-sampling path using deconvolution operations to restore the spatial size of the down-sampled feature maps to the original input dimensions. Inspired by the FCN architecture, the authors in \citep{badrinarayanan2017segnet} proposed encoder-decoder network architecture for semantic pixel-wise segmentation called \textit{SegNet}. Each max-pooling layer in the encoder part has a corresponding up-sampling layer in the decoder, which restores the feature maps to their original resolution. The softmax layer makes pixel-wise predictions and generates the final segmentation output, which has dimensions similar to those of the original input image. Recent successful segmentation networks are mostly based on convolutional encoder-decoder architectures. An illustration of the semantic-wise brain tumor segmentation with a SegNet-like architecture is depicted in Fig. \ref{fig:seg}. 

FCN and its variants have been extended to 3D for voxel classification. There have been several attempts to the application of FCNs in 3D MRI segmentation in the medical literature \citep{brosch2016deep, chang2016fully, shen2017boundary, shen2017efficient}. For instance, a 3D CNN consisting of two interconnected pathways, specifically a convolutional and deconvolutional pathway, was proposed in \citep{brosch2016deep} for multiple sclerosis lesion segmentation in MRI. They connected the first convolution layer and the last deconvolution layer through a single skip connection as a joint learning strategy of the feature extraction and prediction pathways and achieved superior performance when compared with five publicly available methods for this task. The FCN architecture was later modified in \citep{ronneberger2015u}, named U-Net, to improve volume-to-volume learning and yield more precise segmentations, which will be discussed next.

\subsection{Biomedical Image Segmentation with U-Net}

Ronneberger et al. \citep{ronneberger2015u} proposed a U-Net architecture composed of a contracting path and an expanding path, which has become the dominant approach for segmentation in recent applications.  Although similar to an FCN, the architectural novelty in U-Net lies in the fact that the up-sampling and down-sampling layers  are combined by skip connections to connect the opposing convolution and deconvolution layers. Subsequently, several variants of U-Net capable of performing 3D volume-to-volume segmentation were proposed with significant improvements. For instance, in \citep{cciccek20163d}, a 3D U-Net is proposed by replacing 2D operations in a 2D U-Net with their 3D equivalents, and in \citep{milletari2016v} a 3D-variant of the U-Net architecture using residual blocks, called V-net was implemented. These architectures are trained on entire images or large image patches as opposed to small patches and are hence affected by data scarcity, which is typically addressed through data transformations such as shifting, rotating, scaling, or applying random deformations. 

Efficient DNN-based medical image segmentation is often driven by the recent progress in semantic segmentation. In \citep{kayalibay2017cnn}, the authors employed a network architecture similar to 3D U-Net for bone and brain tumor segmentation. They combined multiple segmentation maps created at different scales, which was shown to speed up convergence. However, this approach can be computationally expensive as it employed large receptive fields in convolutional layers. In \citep{isensee2017brain}, the authors implemented a 3D U-Net that has been modified with respect to up-sampling pathways, normalization methods, number of filters, and batch size, enabling training with large image patches and capturing contextual information that leads to segmentation performance improvements. In a more recent work \citep{chen2018s3d} a separable 3D U-Net consisting of separable 3D convolutions was proposed. Their S3D-UNet architecture fully utilizes the 3D volumes via the use of several 3D U-Net blocks. Isensee et al. \citep{isensee2018no} argue that a well-designed U-Net architecture is competent in generating state-of-the-art segmentation maps without the need for significant architectural modifications and that enhancements proposed in prior literature provide no additional benefit. It should be highlighted that the winning contributions of the BRATS challenge in 2019 \citep{jiang2019two} and 2020 \citep{isensee2020nnu} have also adopted U-Net based architectures. The authors in \citep{jiang2019two} utilized a two-stage cascaded U-Net, and Isensee et al. \citep{isensee2020nnu} employed the nnU-Net architecture \citep{isensee2021nnu}, initially developed as a general-purpose network for segmentation following a U-Net. 3D Deep Learning models used in the winning entries of the BRATS challenge within the past five years is summarized in Table \ref{tab:win}.

\begin{table*}[ht!]
\footnotesize
\caption{3D Deep Learning models used in the winning entries of brain tumor segmentation (BRATS) challenge within the past 5 years}
\label{tab:win}
\renewcommand{\arraystretch}{1.5} 
\resizebox{\textwidth}{!}{\begin{tabular}{>{\raggedright\arraybackslash}c p{2.8cm} p{1.8cm} p{8cm}}
\hline
\multicolumn{1}{c}{\textbf{Challenge}} & \textbf{Method} & \multicolumn{1}{>{\centering\arraybackslash}p{1.3cm}}{\textbf{Dice score on test set}}  & \multicolumn{1}{c}{\textbf{Remarks}}  \\ \hline
BRATS 2017 & Ensemble of different architectures: DeepMedic, 3D U-Net,   and 3D FCN \citep{kamnitsas2017ensembles}   & 0.886 (WT) \newline   0.729 (ET) \newline 0.785 (TC)   & Aggregate predictions from an ensemble of multiple   models and architectures. The models in the ensemble are trained and optimized separately under different optimization configurations and loss functions. \\ 
BRATS 2018 & Encoder-decoder based CNN architecture \citep{myronenko20183d}   & 0.8839 (WT) \newline 0.7664 (ET) \newline 0.8154 (TC)& Imaging features are extracted using a larger encoder, and the segmentation mask is reconstructed using a smaller decoder. A variational auto-encoder branch is added for regularization purposes. The ensemble of 10 models yielded the best performance.\\ 
BRATS 2019  & Two-stage cascaded U-Net \citep{jiang2019two}   & 0.8880 (WT) \newline 0.8327 (ET) \newline 0.8370 (TC)   & The first stage is a variant of U-Net that generates a coarse prediction. It is refined in the second stage with a U-Net consisting of two decoders. \\ 
BRATS 2020  & Modified nnU-Net \citep{isensee2020nnu}    &0.8895 (WT) \newline 0.8203 (ET)  \newline 0.8506 (TC) & Incorporate several modifications into the nnU-Net architecture \citep{isensee2021nnu}, including heavy data augmentation,   region-based training, and post-processing.  \\ 
BRATS 2021  & Extended nnU-Net \citep{luu2022extending}   & 0.9319 (WT) \newline 0.8835 (ET) \newline 0.8878 (TC)   & Modifications to the nnU-Net include the use of group normalization, a larger encoder component in the U-Net, and axial attention in the decoder.\\ 
\hline
\end{tabular}}
\end{table*}

Supervised learning methods require large manually annotated data sets and do not generalize well for previously unseen pathological appearances. Alternatively, deep generative models facilitate unsupervised training and provide a mathematically grounded framework for learning hidden structures and complex distributions from unlabeled data. Next, we briefly discuss a subset of deep generative models, generative adversarial networks (GANs) \citep{goodfellow2014generative}. 

\subsection{Generative adversarial networks} 

\begin{figure*}[t!]
    \includegraphics[width=\textwidth]{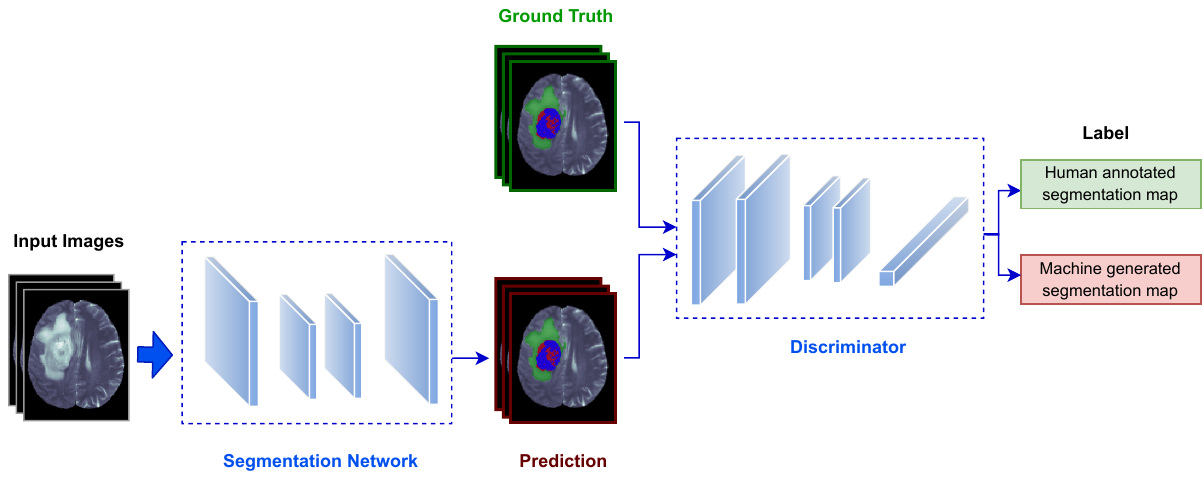}
    \caption{Adversarial training for tumor segmentation. Segmentation network generates segmentation maps and discriminator distinguishes between the manually annotated and predicted segmentation maps.}
    \label{fig:gan}
\end{figure*}

Initially introduced for image synthesis from noise, GANs have had an enormous impact on computer vision. Composed of a generator that generates new data by learning a latent representation of the training data from a prior distribution and a discriminator that discriminates between real and synthetic data, the network is designed to compete against the two components during training, referred to as adversarial training. The same notion can be applied in segmentation by devising a segmentation network in place of the generator such that the discriminator distinguishes the input segmentation mask between the manually annotated ground truth and the segmentation map predicted by the network (Fig. \ref{fig:gan}). This forces the segmentation network to generate segmentation maps that are more anatomically plausible. In \citep{luc2016semantic}, the authors applied an adversarial training approach for semantic segmentation, which is arguably the first study to utilize adversarial training to learn semantic segmentation models. They trained a convolutional net coupled with an adversarial network that can discriminate between a segmentation map from ground truth and model prediction, thus, facilitating the regularization of the segmentation and the results showed improved accuracy. In the case of tumor segmentation, Kamnitsas et al. \citep{kamnitsas2017unsupervised} explored the performance of adversarial networks on domain adaptation and showed that their approach is robust to domain shift and does not require manual annotation of labels when applied to new data in the testing phase. This strategy of employing GANs for brain tumor segmentation was later followed by many other studies \citep{yu20183d, nema2020rescuenet, zhang2021cross} incorporating several modifications. Notable is the \emph{Segan} \citep{xue2018segan} introduced for medical image segmentation, where they show that single scalar output of real or fake in classic GANs is ineffective and proposed an adversarial critic network with a multi-scale $L_1$ loss function.

\begin{table*}[ht!]
\footnotesize
\caption{Brain tumor segmentation studies based on deep learning}
\label{tab:dl}
\renewcommand{\arraystretch}{1.5}
\resizebox{\textwidth}{!}{\begin{tabular}{>{\raggedright\arraybackslash}p{3cm} c c p{8.5cm}}
\hline
\multicolumn{1}{c}{\textbf{Model architecture}} &
\multicolumn{1}{c}{\textbf{Reference}} & 
\textbf{Year} & \multicolumn{1}{c}{\textbf{Remarks}}  \\ 
\hline
Convolutional Neural Networks (CNN)
& \citep{zikic2014segmentation} & 2014 &  2D CNN model architecture trained with a multi-channel patch-based approach\\
& \citep{dvovrak2015local} & 2015 & 2D CNN architecture with multi-modal image patches mapped into four 2D input channels\\
& \citep{pereira2016brain} & 2016 & 2D CNN-based architecture with multi-modality input\\
& \citep{kamnitsas2017ensembles} & 2017 &  Ensemble of multiple models which include 3D CNN-based architectures \\
& \citep{valverde2019one} & 2019 &  3D CNN model trained with multi-sequence 3D image patches\\
Multi-path architectures
& \citep{zhao2016multiscale} & 2016 &   Three-pathway multi-scale 2D CNN architecture \\
& \citep{havaei2017brain} & 2017 &   Patch-based two-pathway 2D CNN architecture \\
& \citep{kamnitsas2017efficient} & 2017 & Dual pathway 3D CNN architecture with 3D fully connected CRF for post-processing\\
& \citep{razzak2018efficient} & 2018 & Multi-scale two-pathway CNN\\
& \citep{sun2021segmentation} & 2021 & Multi-pathway 3D FCN-based solution\\
Fully Convolutional Net (FCN)
& \citep{brosch2016deep} & 2016 & 3D Network consisting of a convolutional and deconvolutional pathways with skip connections\\
& \citep{chang2016fully} & 2016 & Fully Convolutional Deep Residual Neural Network\\
& \citep{shen2017boundary} & 2017 & Boundary-aware FCN based on a multi-task learning approach\\
& \citep{shen2017efficient} & 2017 & Symmetry-driven FCN comprised of a down-sampling path and three up-sampling paths\\
U-Net/V-Net
&  \citep{kayalibay2017cnn} &  2017   & Multiple segmentation maps created at different scales \\ 
& \citep{isensee2017brain} & 2017 & 3D U-Net that has been modified with respect to up-sampling pathways, normalization methods, number of filters, and batch size.\\
& \citep{chen2018s3d} & 2018 & A separable 3D U-Net consisting of separable 3D convolutions.\\
& \citep{jiang2019two} & 2019 & A two-stage cascaded U-Net. The winning contributions of the BRATS challenge in 2019.\\
& \citep{lorenzo2019segmenting} & 2019 & U-Net-based architecture trained with heavy data augmentations.\\
& \citep{li2019novel} & 2019 & An improved U-Net with up skip connections and modified Inception modules. \\
& \citep{isensee2020nnu} & 2020 & Modified nnU-Net \citep{isensee2021nnu} with heavy data augmentations. The winning contributions of the BRATS challenge in 2020.\\
& \citep{hatamizadeh2022swin} & 2022 & U-Net-like architecture design with a Swin transformer as the encoder, called Swin UNETR.\\
Generative Adversarial Networks (GAN)
&  \citep{kamnitsas2017unsupervised} &  2017   & Multi-connected adversarial network proposed for unsupervised domain adaptation\\ 
& \citep{yu20183d} & 2018 & 3D conditional GAN (cGAN) to synthesize FLAIR images from T1\\
& \citep{xue2018segan} & 2018 & GAN-based framewok consisting of a FCN and a adversarial critic network with a multi-scale L1 Loss\\
& \citep{nema2020rescuenet} & 2020 & Residual cyclic unpaired encoder-decoder network with unpaired adversarial training\\
& \citep{zhang2021cross}   & 2021 & CycleGAN-based cross-modality feature learning approach\\
& \citep{wu2021unsupervised} & 2021 &  Unsupervised framework based on symmetric-driven GAN (SD-GAN)\\
\hline
\end{tabular}}
\end{table*}

\subsection{summary} 
In this section, we covered different variants of deep learning architectures ranging from CNNs to GANs and their applications in the segmentation of brain tumors, which are summarised in Table \ref{tab:dl}. Medical image segmentation has seen a massive influx of deep learning methods. Nevertheless, these methods are associated with certain limitations, such as the demand for a large amount of data, the lack of model interpretability and the generalizability \citep{bi2019artificial}. Various strategies have been proposed in past literature to optimize the task of segmentation which has led to significant improvements in the development of fully automated diagnosis systems.

\section{Integrated Statistical/Probabilistic Deep Learning methods for MRI Segmentation}
\label{sec:PDL}

There has been a recent surge of interest in approaches that leverage deep probabilistic models in segmentation. These methods are a step toward bridging the gap between traditional statistical methods and modern deep learning, which enables the implementation of models that significantly benefit from both streams. In the following, we present a comprehensive summary of probabilistic deep learning and the applications of traditional methods combined with deep learning in tumor segmentation.

\subsection{Deep probabilistic models}

Deep probabilistic models combine the virtues of probabilistic modeling with deep learning. These models capture relationships between complex non-linear stochastic structures. For example, variational autoencoder (VAE) \citep{kingma2019introduction} type models are essentially deep statistical/probabilistic learning models that fall at the intersection of statistical and deep learning models, which are also generative Bayesian models but parameterized with a neural network (NN). While there are several other deep probabilistic models, such as deep Gaussian processes \citep{damianou2013deep}, their applications in brain tumor segmentation are lacking in the literature. In this sub-section, we primarily focus on VAE-type models which have been extensively studied in prior literature in the realm of biomedical imaging.

\subsubsection{Variational Auto-encoders}

\begin{figure*}[ht!]
    \centering
    \includegraphics{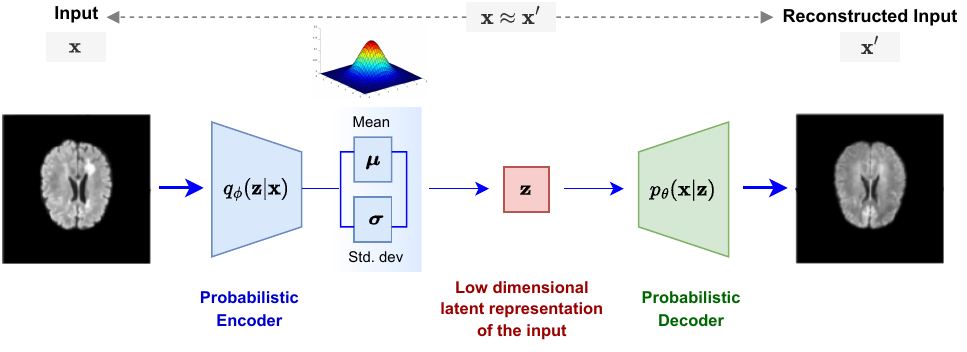}
    \caption{Reconstruction of a healthy brain MRI using a VAE model with Multivariate Gaussian assumption.}
    \label{fig:vae}
\end{figure*}

An autoencoder (AE) is composed of an inference network (encoder) that learns a mapping from the high-dimensional input $\textbf{x} \in \mathcal{X}$ to a lower-dimensional latent representation $\textbf{z} \in \mathcal{Z}$, and a generator network (decoder) that reconstructs the data in latent space to a reconstruction $\textbf{x}^\prime$, where the input space is $\mathcal{X}=\mathbb{R}^D$ and the latent space is $\mathcal{Z}=\mathbb{R}^d$. VAEs provide a probabilistic interpretation for variables in latent space by projecting input data onto a distribution. The prior distribution in the latent representation is typically a multivariate Gaussian, denoted by $p_\theta (\mathbf{z}) = \mathcal{N}(0,\mathbb{I}_d)$. The conditional distribution $p_\theta (\mathbf{x} \, | \, \mathbf{z})$ parametrized by $\theta$ represents the likelihood of generating a true data sample, given the sampled latent vector, which is usually defined as $p_\theta (\mathbf{x} \, | \, \mathbf{z})=\mathcal{N}(\mathbf{x} \, | \, \mu_\theta (\mathbf{z}), \mathbb{I}_D \, \sigma^2_\theta (\mathbf{z}))$ . The corresponding posterior distribution $p_\theta(\mathbf{z} \, | \, \mathbf{x})$ is analytically intractable and hence it is approximated by a variational distribution $q_\phi(\textbf{z} \, | \, \mathbf{x})$, parametrized by $\phi$. In practice, the probabilistic encoder parameterizes the latent distribution and maps input data onto a mean $(\mu)$ and variance $(\sigma)$ vector, which are the parameters of the multivariate Gaussian. Latent distribution is then randomly sampled to feed into the probabilistic decoder that parameterizes the likelihood distribution $p_\theta(\mathbf{x} \, | \, \mathbf{z})$ and projects the latent space attributes back to the original domain space. 

Owing to their ability to model the distribution that characterizes data, these models enable density-based anomaly detection and have received significant attention in clinical imaging. In MRI analysis, VAEs typically address the detection of brain pathology via an unsupervised abnormality detection approach, where the distribution of healthy brain anatomy $\mathbf{x} \in \mathcal{X}_{healthy}$ is modeled, which enables the detection of pathologies as outliers of the distribution. Multiple studies have explored the feasibility of VAEs in tumor segmentation \citep{myronenko20183d,chen2018unsupervised, baur2018deep}. This approach of incorporating VAEs in segmentation architectures is evident in the works of \citep{myronenko20183d}, the winning contribution of BRATS2018 challenge, in which they add a VAE to regularize the decoder. In \citep{chen2018unsupervised}, the authors investigated the performance of VAEs and adversarial auto-encoders on unsupervised lesion segmentation, where they used a prior distribution of healthy brain tissues to detect any abnormalities. They further proposed a latent constraint to improve latent space consistency. Baur et al. \citep{baur2018deep} presented a spatial VAE-based deep generative model that captures the global context of MR slices instead of local patches. A schematic illustration of brain MRI reconstruction with the VAE model based on multivariate Gaussian assumption is presented in Fig. \ref{fig:vae}.

\subsection{Probabilistic neural networks}

Probabilistic neural networks, such as Bayesian neural networks and mixture density networks are deep networks that incorporate probabilistic layers and have many applications, including quantifying uncertainties (sub-section \ref{subsub:uncertain}) in segmentation models. This section primarily reviews the literature on Bayesian neural networks which have been successfully applied in medical image segmentation.

\subsubsection{Deep Bayesian Neural Networks}

In contrast to standard deep neural networks that assign point estimates to the neural network parameters, in the Bayesian probabilistic perspective of deep learning the space of the parameters is specified by prior and likelihood functions. The most probable parameter values can then be captured by the posterior distribution that enables obtaining probabilistic interpretations of model predictions. However, its practical effectiveness is limited by the difficulties arising from choosing an appropriate prior and posterior computation in high-dimensional space. Determining the posterior is analytically intractable, which is typically addressed through an approximation such as variational inference or Markov Chain Monte Carlo (MCMC) based (e.g. Markov Chain Bernoulli dropout \citep{gal2016dropout}) methods. 

In multi-label segmentation of brain tumors, a framework that leverages Bayesian Neural Networks (BNN) and structured random forests in deep learning is reported in \citep{amiri2018bayesian}. The Bayesian network was utilized to encode the statistical relationships between image regions at the super-pixel level. This approach showed superior performance compared to the other methods. In particular, BNNs have proven excellent at modeling uncertainties in bio-medication image segmentation \citep{mcclure2019knowing,chen2022medical}. For example, in a recent work by Chen et al. \citep{chen2022medical}, the authors presented a functional variational BNN where they set the prior and variational posterior distributions as Gaussian Processes. They evaluated its performance on medical image segmentation and showed their approach leads to performance gains on both segmentation and uncertainty estimation tasks.

\subsection{Traditional methods combined with deep learning and their applications in tumor segmentation}
\label{subsec:applications}

Deep and statistical learning based integrated methods have great utility in a multitude of tasks, some of which are briefly discussed here.

\begin{figure*}[ht]
    \centering
    \includegraphics{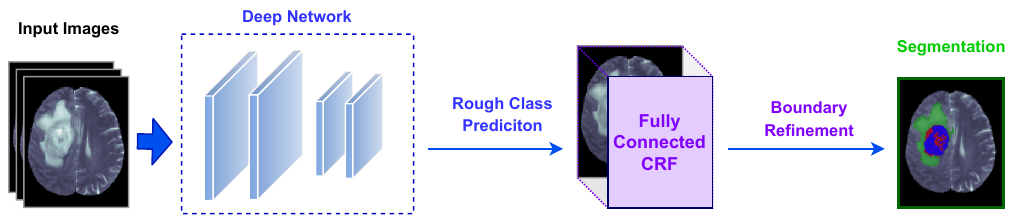}
    \caption{Schematic illustration of post segmentation refinement with a fully connected CRF that operates on the entire segmentation map generated by the deep network to refine coarse voxel-level predictions.}
    \label{fig:crf}
\end{figure*}

\subsubsection{Data set Limitations/ Scarce Annotations}

Probabilistic deep learning has proved to be useful when training with limited data and sparse annotations \citep{dalca2019unsupervised, ito2019semi, kuzina2019bayesian}. The authors in \citep{dalca2019unsupervised} combined probabilistic atlas-based segmentation with deep learning to develop an unsupervised brain MRI segmentation model. They showed that their approach achieves promising results in terms of accuracy without the need for annotated data at test time and is thus computationally efficient. Ito et al. \citep{ito2019semi} modeled the tumor segmentation as a semi-supervised learning problem based on a CNN using a small number of labeled images and a large number of unlabeled images where they use image registration to attach a pseudo-label for every voxel. Since the true labels of the unlabeled images were unknown, they merged a probabilistic framework in their CNN model using the EM algorithm. While E-step performs a maximum a posteriori estimation for the hidden true label of unlabeled images, the M-step updates the parameters of the CNN model based on the estimated true labels. Their experimental results suggested that their approach achieves more accurate and stable segmentation than the models trained relying only on DNN-based methods.

\subsubsection{Segmentation Refinement}

Voxel-based classification methodologies sometimes result in spurious responses and therefore, graphical-based models such as MRFs \citep{shakeri2016sub} and CRFs \citep{kamnitsas2017efficient, lafferty2001conditional, chen2017deeplab, dou20173d, zhao2018deep} are often integrated into likelihood maps generated by CNNs as post-processing to refine the segmentation, which acts as label regularizers (Fig. \ref{fig:crf}). CRFs are probabilistic models that consider the contextual information, which allows to model voxel neighborhoods that are arbitrarily large. More specifically, CRFs model the conditional probability distribution of the target pixel labels given the prediction distribution for each pixel. There are several different variants of post-segmentation refinement strategies with CRF, such as fully-connected CRF (FC-CRF) that operates over the whole segmentation map, locally connected CRF that smoothens local patches, and CRFs formulated as recurrent neural networks (RNN-CRF) \citep{zheng2015conditional} that allow end-to-end training with the segmentation network. In the realm of tumor segmentation, authors \citep{kamnitsas2017efficient} used a 3D CNN model in conjunction with a 3D fully connected CRF for segmenting brain lesions from 3D medical scans. While the 3D CNN model generates soft segmentation maps, CRF acts as a post-processing step that produces final refined segmentation labels. Instead of applying CRFs as a post-processing step for FCNNs, Zhao et al. \citep{zhao2018deep} suggested the integration of FCNNs and CRFs as one deep network. They used 2D image patches and image slices in axial, coronal, and sagittal views and trained three models, which were then combined for the segmentation of brain tumors and their results indicated that FCNN-CRF combination improves segmentation robustness.

\subsubsection{Uncertainty Quantification} 
\label{subsub:uncertain}

There are several other areas, such as uncertainty estimation, where the integration of traditional statistical learning methods with deep learning becomes beneficial. Quantifying deep network prediction uncertainty arising from data set limitations or intrinsic biological noise is often necessary. Prior studies have investigated uncertainty quantification with Bayesian neural networks \citep{kendall2015bayesian, gal2016dropout} and their applicability to biomedical imaging in the context of brain segmentation \citep{jena2019bayesian, roy2019bayesian, kwon2020uncertainty}. In \citep{jena2019bayesian} authors show that uncertainty can be estimated analytically at test time through a Bayesian DL framework. Their experimental results on multiple biomedical imaging data sets indicated improved segmentation and calibration performance. Thus, prior research suggests that employing Bayesian deep nets for brain  segmentation is a promising direction for obtaining uncertainty estimates and the robustness level required for clinically deployed automated systems. 

\subsubsection{Generalizability} 

A model with good generalizability has great clinical potential. The generalizability of a model describes the adaptability of a model when applied to data under different settings or populations not encountered during training. For instance, a model trained with data from one institution should perform equally well on data from another, without significant performance degradation. The same applies to cross-modality segmentation (e.g., CT and MRI). The first federated learning challenge, Federated Tumor Segmentation (FeTS) challenge 2021 \citep{pati2021federated} was introduced recently, in which one of the challenge tasks includes domain generalization. The goal of this task is to assess the robustness of the segmentation model to dataset shifts in multi-institutional glioma MRI scans. Some recent studies \citep{gholami2018novel,agn2019modality, jiang2020psigan, cerri2021contrast, hamghalam2021modality} have indicated that statistical and deep learning hybrid methods can successfully be applied to improve generalizability. For example, Cerri et al. \citep{cerri2021contrast} presented a method for simultaneous segmentation of whole-brain and lesions by utilizing both a probabilistic model and a deep net and demonstrated their approach is contrast adaptive.


\begin{table*}[ht!]
\footnotesize
\caption{Applications of traditional methods combined with deep learning in tumor segmentation}
\label{tab:probdl}
\renewcommand{\arraystretch}{1.5}
\resizebox{\textwidth}{!}{\begin{tabular}{>{\raggedright\arraybackslash}p{3.2cm} p{3cm} c p{8cm}}
\hline
\multicolumn{1}{c}{\textbf{Application}} & \multicolumn{1}{c}{\textbf{Method}} & 
\textbf{Year} & \multicolumn{1}{c}{\textbf{Remarks}}  \\ 
\hline
Segmentation performance
& Encoder-decoder with a VAE branch
\citep{myronenko20183d} & 2018 & VAE branch is added to regualrize decoder. The winning contribution of BRATS2018 challenge.\\
& VAE
\citep{chen2018unsupervised} & 2018 & Unsupervised lesion segmentation.\\                                            
& Spatial VAE-based deep generative model \citep{baur2018deep} & 2018 & Captures the global context of MR slices.\\
& Two-stage cascade model with VAEs\citep{lyu2020two}  & 2020 & VAE was used for regularization to address over-fitting\\
& Context aware network combined with a CRF \citep{liu2021canet} & 2021 & Integrate a feature fusion model based on CRF.\\
Scarce/weak Annotations
& Probabilistic atlas-based segmentation and deep learning \citep{dalca2019unsupervised} & 2019
& Combines probabilistic   atlas-based segmentation with unsupervised deep learning to segment new MRI   scans without requiring the associated labels to train new data. \\                                    
& CNN model combined with EM algorithm \citep{ito2019semi} & 2019 &  Semi-supervised learning framework with an EM algorithm integrated for training with a small number of labeled images.\\
& Generative Bayesian Prior network \citep{kuzina2019bayesian} & 2019 & A knowledge transfer method based on Bayesian generative models for training with small labeled data sets.\\
& Belief Function-Based evidential neural network\citep{huang2021belief}   & 2021 & Based on the belief function theory, where the architecture is composed of a light UNet (LUNet) and an evidential neural network (ENN).\\
Segmentation refinement
& FCNNs and CRFs \citep{zhao2018deep} & 2018 & CRFs are constructed as 
recurrent neural networks. Integration of the CRF-RNN framework into FCNN optimizes the spatial consistency of the predicted brain tumor segmentation map. \\
& CNN architecture with FCRF \citep{chang2019mix} & 2019 &Fully connected CRF is   used as a post-processing method that encode global contextual information in   the prediction generated by the CNN.\\
& Heterogeneous CNNs (HCNN) combined with a CRF-RRNN model \citep{deng2020deep} &
2020 & HCNNs are combined with CRF to increase appearance and spatial accuracy. The authors showed that this procedure also improve segmentation robustness. \\
Uncertainty quantification
& Bayesian Neural Network (BNN) \citep{jena2019bayesian} & 2019 & Uncertainty is estimated analytically at test time through a Bayesian DL framework.\\                                             
& BNN \citep{kwon2020uncertainty} & 2020 & Predictive uncertainty is quantified using a BNN under a variational inference setting. \\                                    
& Functional variational BNN with Gaussian processes \citep{chen2022medical} & 2022 & Variational inference is   performed in function space. The Gaussian process is set as prior and   variation posterior distributions. \\                           
& Region-Based Evidential Deep Learning \citep{li2022region} & 2022 & Output of the neural network is interpreted as evidence which was modelled as a Dirichlet distribution. \\
Generalizability
& Biophysical tumor growth model combined with a GAN \citep{gholami2018novel} & 2018 & Domain adaptation framework\\
& Generative probabilistic model\citep{agn2019modality} & 2019 & Modality-adaptive method in which convolutional restricted Boltzmann machines are used to model the tumor shape. \\
&  Probabilistic segmentation\citep{jiang2020psigan}  & 2020 & Unpaired Cross-Modality Adaptation.\\
&  Generative model combined with a deep net \citep{cerri2021contrast} &  2021   & Simultaneous segmentation of whole-brain and lesions. They demonstrated their approach is contrast adaptive. \\  
& Gaussian process prior VAE \citep{hamghalam2021modality} & 2021 & This approach was proposed to impute missing MRI sub-modalities.\\
\hline
\end{tabular}}
\end{table*}

\subsection{Summary} The research area covered in this section entails probabilistic deep learning and the applications of deep/statistical hybrid methods in medical image segmentation, summarised in Table \ref{tab:probdl}. Probabilistic deep learning has been discussed under two approaches: deep probabilistic models and probabilistic neural networks. The limitations of the current state-of-the-art can be addressed by unifying traditional models with newer deep learning approaches. In summary, it could be argued that the future path for medical decisions with computer-aided systems lies in the merging of traditional statistics with deep learning.

\section{Brain Tumor Segmentation Data sets and Evaluation Metrics}
\label{sec:data}

\subsection{Data sets for brain tumor segmentation}

The performance of segmentation algorithms is highly dependent on the quality of the imaging data used for training. Here we list some of the publicly available data sets for medical image segmentation. 

\begin{itemize}

\item \textbf{Brain tumor segmentation challenge (BRATS)}: The BRATS challenge \citep{menze2014multimodal, bakas2017advancing, bakas2018identifying} is organized annually and the data is published by the University of Pennsylvania's Center for Biomedical Image Computing and Analytics (CBICA). It comprises clinically acquired 3D MRI scans of glioblastoma patients. The BRATS2021 data included pre-operative multi-parametric MRI scans for a total of 2040 cases (1251 for training, 219 for validation, and 570 for testing). The MRI brain scan modalities include T1-weighted, contrast-enhanced T1-weighted, T2-weighted, and T2 FLAIR sequences. Manual annotations by radiologists serve as the ground-truth segmentation labels of tumor sub-regions.  

\item \textbf{Federated tumor segmentation challenge (FETS)}: The FETS is the first challenge in federated learning, which started in 2021 (FETS 2021) \citep{pati2021federated}. In addition to clinically acquired scans from multiple institutions, the FETS 2022 challenge leverages multi-institutional, multi-parametric MRI (mpMRI) scans from the BRATS 2021 challenge. The challenge focuses on developing models for segmenting intrinsically heterogeneous gliomas. The main goals include developing a consensus model using information from data collected by various independent institutions but without data sharing and evaluating the developed models under a federated setting.

\item \textbf{Medical segmentation decathlon (MSD)}: MSD is a biomedical image analysis challenge \citep{simpson2019large,antonelli2022medical} that focuses on a multitude of clinical tasks. It contains large, open-source medical imaging data sets that serve as a benchmark for validating and testing the generalizability of image analysis algorithms. Its brain tumor data set comprises multi-modal (FLAIR, T1w, T1gd, T2w) MRI scans of patients diagnosed with glioma. Data is acquired from multiple clinical sites and includes 4D volumetric MRI scans for a cohort of 750 patients (484 for training and 266 for testing).


\end{itemize}

Other data sets in brain lesion detection include the Ischemic Stroke Lesion Segmentation challenge (ISLES 2022) \citep{petzsche2022isles} data, Brain Tumor data set from Figshare \citep{cheng_2017},  and The Cancer Genome Atlas (TCGA) data sets: TCGA-GBM \citep{scarpacel.mikkelsen2016, clark2013cancer, bakas2017advancing} and TCGA-LBM \citep{pedano_2016, clark2013cancer, bakas2017advancing}.

\subsection{Evaluation metrics in medical image segmentation}

Assessing the quality of segmentation plays an important role in model development, where model predictions are compared against manual annotations by a radiologist, considered the ground-truth label. Segmentation performance can be evaluated under several metrics, some of which are discussed below. In the following metrics, the ground truth segmentation is represented by $S_g$, and the segmentation that is evaluated is denoted by $S_p$.

\begin{itemize}

\item \textbf{Dice coefficient}: The Dice coefficient, the most widely used metric for evaluating segmentation performance, measures the spatial overlap between the ground truth $(S_g)$ and predicted $(S_p)$ segmentations. Its value range is between 0 and 1, with a dice coefficient equal to 1 indicating a complete overlap. Mathematically, it is defined as follows:

\begin{equation}
    Dice (S_g,S_p) = 2 \frac{|S_g \cup S_p|}{|S_g|+|S_p|}
\end{equation}

\item \textbf{Jaccard Index}: The Jaccard index, also known as the intersection over union metric, is related to the dice coefficient. It is computed by dividing the intersection of the two sets by their union: 

\begin{equation}
    Jaccard (S_g,S_p) = \frac{|S_g \cap S_p|}{|S_g \cup S_p|}
\end{equation}

\item \textbf{Hausdorff distance}: 

Hausdorff distance $(HD)$ is a spatial distance-based metric that quantifies the distance between the ground truth and segmentation boundaries. Let $i$ and $j$ denote the points in $S_g$ and $S_p$, respectively. Let the distance between the points $i$ and $j$ be denoted by $d$. The $HD$ can be calculated as follows:

\begin{equation}
    HD = \left( \max_{i \in seg} \left( \min_{j \in {gt}} (d(i,j)) \right), \max_{j \in {gt}} \left( \min_{i \in {seg}} (d(i,j)) \right) \right) 
\end{equation}

\end{itemize}

Other metrics for segmentation evaluation include \emph{sensitivity} and \emph{specificity}. These measures, however, have drawbacks, such as small segments being penalized more than large segments \citep{taha2015metrics}. We have only discussed the most commonly used metrics here. The interested reader is referred to \citep{taha2015metrics} for a comprehensive overview of metrics for evaluating the segmentation of medical images.

\color{black}

\section{ Limitations, challenges and future directions in biomedical image analysis}
\label{sec:limitations}

\subsection{Limitations and challenges}

Medical image segmentation encompasses a broad set of challenges including complex and subtle surrounding boundaries of organs, lack of sufficient level of the region uniformity and similarity, low contrast and intensity inhomogeneity, image noise, partial volume effect, and other artifacts that impede precise identification of abrupt variations between organs of interest. Class imbalance, data acquisition, and the 3D nature of data pose additional challenges, which are briefly described here.

\begin{itemize}

\item \textbf{Imbalanced Data}: \emph{Class imbalance} presents one of the major difficulties in segmentation. For example, the tumor region typically constitutes only up to 15\% of the brain, and tumor sub-regions are even smaller, thus the class distribution is dominated by the background voxels compared to the foreground voxels of tumor tissues. Prior work propose several strategies to address imbalanced distributions, which include applying a weight map to the categorical cross-entropy loss function or incorporating different loss functions, such as focal loss \citep{lin2017focal}. A recently introduced dynamically weighted balanced loss \citep{fernando2021dynamically} focuses on dynamically adjusting class weights to address the class imbalance that also leads to improved calibration. In the context of tumor segmentation, customized dice loss is a commonly adopted approach \citep{milletari2016v, sudre2017generalised}. Devising algorithms that are adept at dealing with class imbalance is imperative in medical image analysis.

\item \textbf{Scarce Annotations}: Another primary limitation in biomedical imaging is the \emph{acquisition of labeled data}. While computer-aided systems make the diagnosis process more efficient, supervised deep architectures require access to large, labeled data sets. Manual annotation requires domain expertise, and it is time-consuming. Since small data sets lack generalizability, deployment of semi or unsupervised architectures may be more effective. 

\item \textbf{Memory Consumption}: \emph{3D neuroimaging data} are memory intensive. In order to utilize contextual information in brain volumetry on MR images, 3D deep networks are typically deployed, which use extensive memory. Previous research adopts different techniques to address this issue, which include training on 2D slices or image patches and employing multi-path models. However, regulating the trade-off between leveraging the spatial context of neighboring voxels and computational complexity continues to be a challenge. Although some efforts have been made to address this issue \citep{chen2016training, roth2018application, reina2020systematic}, further research is required in this area.
\end{itemize}

\subsection{Future Directions}

The transition from scientific research to clinical practice depends on the trustworthiness of computer-aided diagnosis systems and raises profound questions about reliability, uncertainty, and robustness. For example, how to detect when the computational algorithms get it wrong? Can we predict when the machine learning model will fail and how to make the algorithm robust to changes in real-world conditions and clinical data? These research questions are of central interest in safety-critical medical applications where accuracy is integral. Here we briefly discuss some of the critical challenges and potential future directions in medical image analysis.

\begin{itemize}
\item \textbf{Learning the right features}: The performance of the diagnosis system should not be influenced by the variability across different scanners, settings of the image acquisition, or other clinical conditions of patients. Learning discriminative features, which are generalizable representations across data sets is an active area of research, however, until substantial progress is made in robustness research, caution should be exercised in relying on these strategies where precision is vital. 

\item \textbf{Knowing when the model does not know}: Quantifying uncertainty in tumor segmentation is indispensable in clinical decision-making \citep{begoli2019need, jungo2019assessing}. While deep learning-based approaches have achieved remarkable success performance-wise, their practical applicability is limited as models do not produce a measure for prediction uncertainty. We expect the research on uncertainty estimation in neuroimaging to increase in the near future.

\item \textbf{Detecting when an automated system fails}: Deep network based methods are prone to overfitting and hence evaluating the level of accuracy on new clinical data in the absence of ground truth is crucial in automated diagnosis. Methods are required to identify when the system fails and to evaluate the actual performance after implementation when a reference segmentation by an expert is inaccessible. This has been previously evaluated only to a very limited extent \citep{valindria2017reverse}. Future efforts should further explore this issue before the clinical adoption of automated systems.

\item \textbf{Going beyond human-level performance}: The ultimate goal in computer-aided medical screening is to surpass human-level performance. Cancer histopathology reads in biomedical imaging require expert knowledge, and it may be subject to annotation bias while leading to a lack of agreement among human experts. Therefore, identifying the exact ground truth segmentation could be challenging. Synthesis of ground truth via adversarial learning appears to be a promising research direction in this context. 
\end{itemize}

\section{Conclusion}
\label{sec:conclude}

Precise tumor delineation is imperative for accurate assessment of the relative volume of lesion sub-components. A comprehensive review is conducted on the state-of-the-art brain tumor segmentation of 3D MRI ranging from traditional statistical approaches to modern deep learning. We covered the fundamentals, discussed the methodological foundations, and addressed their potential and limitations. The review conducted leads to the following conclusions: deep learning, being capable of learning high-level feature representations automatically, has emerged as a powerful mechanism in automated volumetric medical image segmentation. Moreover, it has been observed that engrafting statistical methodologies in deep networks improve biomedical image segmentation. Thus, volumetric 3D MRI segmentation at the intersection of statistical modeling and deep learning can be considered a more effective alternative to the methods which are purely statistical or deep learning based. While these automated methodologies improve the qualitative expertise of clinicians and facilitate accurate volumetric tumor delineation and outcome prediction, they mostly remain in the pre-clinical research domain. Although studies have been conducted by many researchers for 3D volumetric tumor segmentation, difficulties associated with biomedical imaging are still insufficiently studied and the deployment of robust automated systems is a challenge for future research to explore. Future studies could fruitfully investigate these issues further through more systematic and theoretical analysis.


\begingroup
\setstretch{0.5}
\setlength{\bibsep}{5pt}
\def\bibfont{\small}


\bibliographystyle{elsarticle-num.bst}
\bibliography{refs}

\endgroup

\end{document}